\pdfoutput=1
\newcommand{\comment}[1]{}

\documentclass{article}
\usepackage{spconf,amsmath,graphicx,url,times,booktabs, tabularx}
\usepackage{multirow}
\usepackage{microtype}
\usepackage{siunitx,makecell,arydshln}
\usepackage[table]{xcolor}

\title{Towards high resolution weather monitoring \\ with sound data}

\name{Enis Berk \c{C}oban$^{1}$ \qquad Megan Perra$^{2}$ \qquad Michael I Mandel$^{1}$}
\address{$^1$ The Graduate Center, CUNY, New York, NY, USA,
  \\ 
  $^2$ Environmental Biology, SUNY-ESF, Syracuse, NY, USA,
  \\
}

\begin{document}
\ninept

\maketitle

\begin{abstract}
Across various research domains, remotely-sensed weather products are valuable for answering many scientific questions; however, their temporal and spatial resolutions are often too coarse to answer many questions. 
For instance, in wildlife research, it's crucial to have fine-scaled, highly localized weather observations when studying animal movement and behavior.  
This paper harnesses acoustic data to identify variations in rain, wind and air temperature at different thresholds, with rain being the most successfully predicted.  
Training a model solely on acoustic data yields optimal results, but it demands labor-intensive sample labeling. Meanwhile, hourly satellite data from the MERRA-2 system, though sufficient for certain tasks, produced predictions that were notably less accurate in predict these acoustic labels.
We find that acoustic classifiers can be trained from the MERRA-2 data that are more accurate than the raw MERRA-2 data itself.  By using MERRA-2 to roughly identify rain in the acoustic data, we were able to produce a functional model without using human-validated labels.  
Since MERRA-2 has global coverage, our method offers a practical way to train rain models using acoustic datasets around the world. 
\end{abstract}


\section{Introduction}\label{sec:intro}

Fine scale weather data is difficult to obtain in the absence of weather stations.
Recording audio data in remote regions is an affordable and scalable solution to the challenge of maintaining weather stations in remote regions, as vast arrays of acoustic recording units can be deployed over large areas for several months or years at a time with little maintenance or intervention. 
These recorders collect `soundscape' information, and a major component of that soundscape is `geophony', i.e., sounds from abiotic sources like wind or rain
\cite{Pijanowski2011}. 
Rain and wind, in particular, produce sounds that are easy for the human ear to classify. Previous research has leveraged existing acoustic indices or power spectral density to automate rainfall detection in acoustic data \cite{Ferroudj2014,BEDOYA2017}. 
Convolutional neural networks (CNN) have also been used for automated rain detection \cite{Avanzato2020} trained from direct measurements in the same location.  Here we present a novel model that can recognize multiple weather variables--rain, wind, and air temperature--at a fine scale.  It is the first example of coarse satellite data (with sampling periods of an hour or more) being used with unlabeled acoustic data to produce fine-scale (sub-minute) predictions.  This is a significant leap in weather modeling and model training efficiency because it requires little manual labeling and produces outputs at a scale finer than most remote-sensing products.

Supervised machine learning for rain and wind prediction is resource-intensive due to the necessity of labeling a large volume of samples \cite{Ferroudj2014, wang2022rainfall}. 
Additionally, threshold-based methods using Power Spectrum Density and Signal-to-Noise Ratio thresholds to detect heavy rainfall, require less labeled data but underperform in less rainy or temperate climates, owing to insufficient sound amplification from rain and sparse vegetation \cite{metcalf2020hardrain}.
This study is the first of its kind to incorporate satellite weather data (i.e., MERRA-2) into the sound recognition of weather events. Using this data for training is important for two main reasons: First, it is readily available to the general public and second, it does not have the bias of the sound signatures of specific recording equipment or sampling locations like training data that has been manually labeled. For validation, we manually labeled a relatively small subset of samples, rendering our methodology adaptable to a wide range of environmental conditions.

Nearby weather stations or rain gauges have been used in attempts to automate the labeling process\cite{Ferroudj2014, Avanzato2020, wang2022rainfall, chen2022estimating}. However, this may result in less diverse datasets and does not translate to all ecosystems, especially those in remote regions---like those in the Arctic---where weather stations are sparse and maintaining rain gauges imposes logistical challenges and additional costs.
This approach is also impractical for small or mobile acoustic devices, such as audiologgers attached to animals \cite{Wijers2018, Latorre2020}, where obtaining measurements for multiple weather variables at each location is critical for animal movement ecology and was previously only feasible through remote-sensing~\cite{Neumann2015}.

This paper trains acoustic models using weather information from MERRA-2 \cite{merra2} with a granularity that approximates a grid wherein each cell spans an area of around 50 km by 50 km with a temporal resolution of 1 hour. However, for the purposes of this study, the MERRA-2 data was spatially downscaled via MicroMet~\cite{micromet}, a meteorological distribution system, to a higher resolution of 250 meters by 250 meters.
While this level of detail is useful for understanding general weather conditions, the study of meteorological impacts on wildlife foraging, migration, and survival sometimes necessitates finer-grained data.  For example: wind affects bird flight paths and energetics \cite{Shepard2016}, air temperatures affects insect activity \cite{Witter2012}, and rainfall and humidity drive foraging and vocalization patterns in bats \cite{Chaverri2017, Appel2019}, all at a finer temporal or spatial scale than most weather products.

Measuring these variables together can help identify severe weather events, such as `rain-on-snow' (ROS).  ROS events are difficult to detect reliably by satellite in mid winter \cite{Serreze2021}, and may have profound effects on wildlife survival and inter-annual population patterns \cite{Forbes2016, Hansen2013}.  
High-resolution, real-time weather information could revolutionize the understanding of the effects of weather variables on wildlife.

\section{ DATA AND MODEL ARCHITECTURE}\label{sec:DATA_MODEL}

\subsection{Weather Data}\label{sec:data:weather_data}
In this study, we utilize weather data obtained from the NASA atmospheric
reanalysis dataset called the Modern-Era Retrospective Analysis for Research
and Applications, Version 2 (MERRA-2) \cite{merra2}.
MERRA-2 features an extensive selection of atmospheric parameters, such as
temperature, precipitation, wind, humidity, and numerous others. These
parameters can be found at different spatial and temporal resolutions, varying
from hourly to monthly timeframes and covering scales from global to regional.

Our partners used MicroMet~\cite{micromet}, a meteorological distribution system, for spatial downscaling of weather data sourced from MERRA-2.
The system accounts for spatial variability in terrain-influenced climates, providing results with horizontal grid increments increments of 250 meters by 250 meters.
It is worth noting that MERRA-2 has a native resolution of 0.5 degrees latitude by 0.625 degrees longitude.
In the context of our study, which is situated at latitudes between 64 and 70 degrees, this equates to approximately 50 kilometers for 0.5 degrees of latitude and between 24 and 30 kilometers for 0.625 degrees of longitude.

\subsection{Sound Datasets and Data Processing}
Original audio recordings were acquired using Acoustic Recording Units (ARUs) from the Arctic and boreal regions, specifically in northern Alaska and Yukon Territories, at latitudes 64$^{\circ}$--70$^{\circ}$ N and longitudes 139$^{\circ}$--150$^{\circ}$ W. This range spans across the Arctic Coastal Plain and includes diverse ecosystems such as tundra, shrub, and boreal forests.
    
A total of 70 ARUs, model SM4 by Wildlife Acoustics, were deployed across the Alaskan and Yukon North Slope, and into the boreal forest.
The ARUs recorded audio at a 48 kHz sampling rate with gains set to 16 dB. Each unit captured 150 minutes of audio per session, with rotating breaks between 120 to 150 minutes. This allowed for a 24-hour cycle coverage over a 4-day period. From these recordings, 10-second clips were extracted for analysis.

This study employs two datasets, both originating from the same raw audio recordings but labeled differently: one using human annotations, and the other utilizing weather data from the MERRA-2 dataset.
The human-annotated dataset is a subset of the EDANSA-2019 dataset, an extensive ecoacoustic collection that we previously compiled and made available in a separate publication \cite{Coban2022}. 
Random 10-second clips were extracted from across the recording season.
An expert meticulously analyzed these clips, studying spectrograms to identify and label all foreground sound. The labels from the EDANSA-2019 dataset utilized in this study were \textit{Wind} and \textit{Rain}.

The second dataset, referred to as the ``satellite training set'' utilized solely for the purpose of training the models. In constructing this dataset, 15,000 ten-second audio clips were randomly sampled across all recording sites and times.
Samples from both satellite and EDANSA datasets were synchronized with corresponding meteorological variables such as rainfall and wind speed from the MERRA-2 dataset. It is worth noting that there were instances where a single one-hour time window from the MERRA-2 dataset was matched with as many as 20 of our ten-second audio clips.
To ensure our models are able to generalize across recording locations, data from each recording site was exclusively included in one of the training, validation, or test sets.

Some data were rendered unusable due to ``clipping'', a distortion that occurs when the sound amplitude exceeds the recording device's dynamic range. We designated samples as ``clipped'' if they reached the maximum or minimum integer value. 
In both the EDANSA and satellite training set, we filtered out 10~s clips where 20\% or more of the samples were clipped.

\subsection{Model Architecture}\label{sec:method:modelarchitecture}

In this study, we aimed to evaluate whether variables derived from weather models such as rainfall, wind speed, humidity, and air temperature could be discerned or estimated through the analysis of sound data.
To assess how these theoretical relationships are reflected in real-world data, we employed a CNN architecture  \cite{Piczak2015-sed-cnn}. The inputs to our model are log-scale mel frequency spectrograms extracted from 10-second audio clips. Specifically, we compute log-scale mel frequency spectrograms with a window size of 42 ms, a hop size of 23 ms, and 128 mel frequency bins. The CNN comprises four convolutional layers with a kernel size of $5 \times 5$. After these convolutional layers, global temporal pooling is applied to reduce the spatial dimensions of the feature maps. The pooled feature maps are then fed into two fully connected layers. Data augmentation techniques, such as SpecAugment and Gaussian noise injection, were integrated to improve the model's performance.

The training process is designed to span 1500 epochs, but could conclude earlier if the validation set performance does not improve over 20 consecutive epochs.
For classification experiments, the model with the highest F-1 score on the validation set is selected, whereas, for regression experiments, the model with the lowest Mean Squared Error (MSE) on the validation set is chosen. The selected model is then used for performance evaluation on the test set.

We compare two approaches to model training, creating individual models for each variable and a shared model to predict all variables simultaneously. This strategy aims to harness the advantages of both specialized and generalized models. The individual models are tailored to capture the unique characteristics and patterns of each variable, potentially enhancing prediction accuracy. Conversely, the shared model, trained on all variables together, is designed to identify and exploit the interdependencies and correlations between the variables, which may be overlooked by the individual models.

\section{Regression Approach}
\label{sec:method:regression}

\begin{table*}[t]
\centering
\begin{tabular}{
  l
  c
  c
  c
  c
  l
  c
  c}\toprule
\textbf{Label} & \textbf{Min} & \textbf{Median} & \textbf{Max} & \textbf{Baseline RMSE} & \textbf{Exp} & \textbf{RMSE} & \textbf{Change (\%)} \\
\midrule
\multirow{2}{*}{Rainfall (mm/hr)} & \multirow{2}{*}{0} & \multirow{2}{*}{0.009} & \multirow{2}{*}{2} & \multirow{2}{*}{0.145} & Individual & 0.145 & -0.233 \\
                 & & & & & Shared & \textbf{0.137} & \textbf{-5.87} \\
\midrule
\multirow{2}{*}{Humidity (\%)} & \multirow{2}{*}{36.5} & \multirow{2}{*}{86.9} & \multirow{2}{*}{100} & \multirow{2}{*}{13.1} & Individual & 11.6 & -11 \\
              & & & & & Shared & \textbf{10.3} & \textbf{-21.5} \\
\midrule
\multirow{2}{*}{Wind Speed (m/s)} & \multirow{2}{*}{0.1} & \multirow{2}{*}{2.39} & \multirow{2}{*}{10.7} & \multirow{2}{*}{1.55} & Individual & 1.18 & -23.6 \\
                 & & & & & Shared & \textbf{1.17} & \textbf{-24.7} \\
\midrule
\multirow{2}{*}{Temperature $^{\circ}$C} & \multirow{2}{*}{-32.3} & \multirow{2}{*}{7.03} & \multirow{2}{*}{22.8} & \multirow{2}{*}{8.34} & Individual & \textbf{5.88} & \textbf{-29.5} \\
                        & & & & & Shared & 6.22 & -25.4 \\
\bottomrule
\end{tabular}
\caption{
Regression results: RMSE comparison on EDANSA's validation set for meteorological variables. It includes EDANSA training set's data distribution (Min, Median, Max), baseline and model RMSE, and change (\%). Negative change signifies better model performance. 'Individual' and 'Shared' refer to models trained for each variable separately and a single model predicting all variables, respectively.}
    \label{table:mse_comparison}
\end{table*}

The MERRA-2 dataset records the four meteorological variables -- rainfall, wind speed, relative humidity, and air temperature -- as continuous values. Specifically, rainfall is recorded in millimeters per hour, temperature in degrees Celsius, relative humidity as a percentage, and wind speed in meters per second. Consequently,datasets, which we utilize in regression experiments, contains these variables in their respective units.

Our models were specifically trained on the satellite training set. Subsequently, we evaluated the performance of our models using the EDANSA validation set. 
To establish a performance baseline, we employ a simple constant prediction strategy, where the model always predicts a single fixed value, irrespective of the input. This fixed value is the mean of the training set of EDANSA data. We compute the MSE of these constant predictions on the EDANSA validation set, and use this MSE as our baseline.

Table \ref{table:mse_comparison} contrasts the square root of MSE (RMSE) losses for each variable between the baseline and the model. Performance change is quantified as the percentage change in RMSE relative to the baseline, with negative changes indicating better performance. 
The model losses for Rainfall is marginally lower than the baseline losses, with percentage changes of $-$5.87\%.
This slight improvement suggests that the models' predictive capabilities for this variable are nearly equivalent to a strategy based on the mean of the training set, indicating that the models may not be effectively capturing the inherent variability in rainfall.
In contrast, the models substantially reduce loss for Humidity, Temperature and Wind Speed by approximately 21, 29\% and 25\% in RMSE, respectively.
This indicates that the models are successful in extracting meaningful information from the audio to predict these variables.

\begin{table}
\centering
\begin{tabular}{ccccc}
\toprule
\multirow{2}{*}{Frequency (Hz)} & \multirow{2}{*}{R. Humidity (\%)} & \multicolumn{2}{c}{Temperature ($^\circ$C)} \\
\cmidrule{3-4}
& & 0.3 & 9.3 \\
\midrule
\multirow{2}{*}{125} & 77 & 0.38 & 0.39 \\
& 91 & 0.36 & 0.35 \\
\midrule
\multirow{2}{*}{4000} & 77 & 50.30 & 31.08 \\
& 91 & 42.35 & 26.46 \\
\bottomrule
\end{tabular}
\caption{Attenuation(dB/km) of sound in air for combinations of the 25th and 75th percentiles of relative humidity and temperature observed in our dataset and for different frequencies.
}
\label{tab:combined_absorption}
\end{table}

Our results reveal a dichotomy in the performance of the individual models versus the shared model. For Rainfall, Relative Humidity and , Wind Speed the variables with worse results, the shared model outperforms the individual models, albeit mostly marginally. This suggests that the shared model can leverage the interdependencies between these variables to achieve a slight improvement in prediction accuracy. Conversely, for Temperature, the variable with best results, the individual model perform better than the shared model. This indicates that the individual models are more successful in capturing the unique characteristics of this variable, leading to a significant improvement in prediction accuracy.

One way that humidity and air temperature might be predictable in the audio recordings would be because of the changes that they induce in the attenuation of sound. When sound travels through the air, some of its energy is converted into heat. This happens due to factors like heat conduction and shear viscosity, among others. These factors are especially noticeable at high frequencies and over long distances, making the air act as a low-pass filter. Both temperature and humidity can affect these processes, and their effects are most noticeable at certain frequencies \cite{attenborough2014sound}. Using formulas from \cite{harris1966absorption}, the effect of these two variables on the attenuation of sound is examined in Table~\ref{tab:combined_absorption} when each variable is set to the 25th and 75th percentiles of its value across our dataset.

In summary, the regression experiments reveal that audio recordings contain enough information for making inferences about rain precipitation, air temperature, humidity, and wind speed. The choice between using individual models or a shared model depends on the specific variable being predicted. Individual models exhibit superior performance for temperature and wind speed, while the shared model has a slight advantage for rainfall and relative humidity.

\section{Classification Approach}
\label{sec:method:classification}
In this section, we adopt a classification approach to analyze meteorological variables. Given the dynamic nature of weather events such as rain, which can evolve within an hour, assigning a singular continuous value to all 10-second clips in an hour could be insufficient to capture the full complexity of the situation.

After conducting regression experiments in section \ref{sec:method:regression}, we observed that audio data contain valuable information that is useful for making predictions about rain precipitation, air temperature, and wind speed. However, the predictions were not as accurate and the error margins were not as low as desired for practical applications.
To achieve higher accuracy and lower error margins, the approach was modified to address this as a classification problem, converting continuous weather variables into binary classes based on thresholds. These thresholds aim to represent the presence or absence of weather events under certain conditions.

The Binary Cross-Entropy loss function was employed for this classification task. 
Our models utilized MERRA-2 data, which was spatially downscaled via MicroMet to higher resolutions  (250 meters $\times$ 250 meters), as detailed in Section \ref{sec:data:weather_data}. This downscaling, though improving resolution, may introduce inaccuracies.
To mitigate this, we employed a hybrid approach that also incorporated the EDANSA dataset which contains manual annotations that have undergone rigorous verification, ensuring higher reliability. 
Our models were specifically trained using the satellite training set. We then report the validation and test results against the Rain and Wind classes from the EDANSA labels. This approach allowed us to leverage both the breadth of satellite and the precision of EDANSA. 

To establish a baseline for comparison, we conducted a similar training and testing process as we did with the satellite and EDANSA datasets. We matched EDANSA samples with their corresponding continuous weather values from MERRA-2. Then, to predict the Rain label using continuous values, we applied thresholds ranging from 0 to the maximum weather variable value with 0.001 increments on the training set. We selected the threshold that yielded the best F1 score on the training set. We used this threshold on the test set of EDANSA to calculate the baseline score.

In our study, we established specific thresholds for each meteorological variable based on the characteristics of the satellite training set and the performance of our baseline experiment. For rainfall, we chose thresholds of 0, 0.06, and 0.1 mm/hr. The 0.06 mm/hr threshold corresponds to the mean of the satellite training set, while 0 mm/hr was chosen as it would be the perfect threshold for rain detection if the MERRA data was flawless. The threshold of 0.1 mm/hr was selected as it yielded the best F1 score on the training set in the Satellite baseline experiment. 
Similarly, for wind speed, thresholds were set at 0.361, 1.1 and 2.471 m/s. These values mark significant shifts in the wind patterns, with 1.1 m/s being the first quartile and 2.471 m/s the mean of the satellite training set and 0.361 m/s being the threshold that yielded the best F1 score in the Wind baseline experiment.

However, it's worth noting that the EDANSA dataset does not contain labels for the temperature and humidity classes since these are not qualities that humans can hear. 
As such, we only consider Rain and Wind classes for the classification experiments.

\begin{table}
\centering
\setlength{\tabcolsep}{5pt}
\begin{tabular}{|c|c|c|c|c|c|}
\toprule
\multirow{2}{*}{Class} & \multirow{2}{*}{Training} & \multirow{2}{*}{Model} & \multirow{2}{*}{Threshold} & \multirow{2}{*}{AUC} & \multirow{2}{*}{F1} \\
& & & & & \\
\midrule
\multirow{4}{*}{Rain} & Satellite & Baseline & 0.1 mm/hr & 0.858 & 0.569 \\
& Satellite & Shared & 0.1  mm/hr  & 0.860 & 0.623 \\
& Satellite & Individual & 0.1  mm/hr  & \textbf{0.911} & \textbf{0.712} \\
& EDANSA & Individual & N/A & 0.969 & 0.848 \\
\midrule
\multirow{4}{*}{Wind} & Satellite & Baseline & 0.361 m/s & 0.659 & 0.632 \\
& Satellite & Shared & 2.471 m/s & 0.815 & 0.766 \\
& Satellite & Individual & 2.471 m/s & \textbf{0.869} & \textbf{0.791} \\
& EDANSA & Individual & N/A & 0.932 & 0.855 \\
\bottomrule
\end{tabular}
\caption{Classification results on the test set of the EDANSA dataset. The ``Threshold" column displays values used to convert continuous weather variables into binary class labels. The ``Model" column represents different model types or training styles: ``Individual" signifies a model trained specifically for a single class, ``Shared" indicates a single model trained for all four variables (Rain, Wind, Temperature, Humidity),
and ``Baseline" denotes the use of Merra-2 data to predict EDANSA labels without any model training involvement.}
\label{table:satellite2edansa_results}
\end{table}

\subsection{Results and discussion}
\label{sec:results}

The performance of our models in classifying weather events based on audio data is comprehensively summarized in Table \ref{table:satellite2edansa_results}. It illustrates the classification results on the test sets of the EDANSA across various thresholds for different meteorological variables, namely rainfall and wind speed. Performance metrics including the Area Under the Receiver Operating Characteristic Curve (AUC), and F1 score, are used for evaluation.

In this table, the ``Threshold" column indicates the value used to convert the continuous weather variables from the satellite training set into binary class labels. An absence of a threshold implies that the training samples originated from the EDANSA dataset (i.e., were created by human listeners based on the audio). The ``Model" column represents different model types or training methodologies: ``Individual" denotes a model specifically trained for a single class, ``Shared" refers to a single model trained for all four variables (Rain, Wind, Temperature, Humidity), 
and ``Baseline" represents the use of Merra-2 data to predict EDANSA labels without any model training.

The results for Rain and Wind in Table \ref{table:satellite2edansa_results} were obtained using the test set of the EDANSA dataset.  It should be noted that, except for the models labeled as EDANSA, all others were trained on the satellite training set.

Our experiments were designed to evaluate the effectiveness of using globally available low-resolution data to train high-resolution audio classifiers.
The EDANSA dataset is biased towards levels of activity detectable by human hearing. Therefore, the model's success not only indicates its performance in lower resolution predictions but also its alignment with human perception of sound activity.

We expected that a single model trained for all variables might perform as well as individually trained models, assuming adequate capacity. 
Individual models performed approximately 6\% better than shared models for both Rain and Wind classes. This discrepancy might be due to the phenomenon of regression to the mean, where the underperformance of certain categories could be dragging down the overall performance, much like the superior categories previously elevated the underperforming ones.

The model's performance on the EDANSA validation set improved with higher thresholds. For rainfall, an AUC of 0.858 and an F-1 of 0.607 at 0 mm/hr rose significantly to 0.937 and 0.735 respectively at 0.1 mm/hr. Similarly, for wind speed, the AUC and F-1 scores improved from 0.799 and 0.736 at 1.1 m/s to 0.839 and 0.768 at 2.471 m/s. However, a 0.361 m/s threshold was not modeled due to a lack of sufficient data below this threshold.
In the table~\ref{table:satellite2edansa_results}, we report only the model with the highest threshold for each variable, as these models demonstrated the best scores on the validation set.

The variance in model performance across different thresholds could be attributed to three potential factors: a) our human labeler possibly missing low precipitation rain samples, b) a higher error rate within the Satellite training set for lower precipitation levels (MERRA-2 can be prone to over predicting light rain events in the Arctic summer months (\cite{Barrett2020})), and c) the transient nature of low precipitation rain events, which could be missed during the conversion from an hourly to a ten-second granularity.

The final row under the ``Rain" and ``Wind" class shows the model's performance when both training and testing were conducted on the EDANSA dataset. Although the performance was superior compared to training on Satellite data, this approach requires manually labeling many samples, which is a time-consuming process.

\section{Conclusions and future work}

This study showed that combining audio with labels from weather satellites can provide higher resolution predictions without costly human annotation. By training our models on satellite data, which was free of equipment or location bias, and validating them with high-quality annotations from the EDANSA dataset, we ensured that our approach is both robust and widely applicable. This methodology proved especially beneficial in recognizing rain and wind events.

This research opens new avenues in the utilization of sound for monitoring weather events, especially in applications where high temporal resolution and localized data are crucial.

Future research can explore a variety of approaches including active learning for dataset annotation, transfer learning for applications such as animal species detection, integration of audio data with satellite information, and the use of student-teacher methodologies for model refinement. These approaches have the potential to enhance weather predictions and ecological monitoring by leveraging sound data with meteorological information.

\section{ACKNOWLEDGMENT}\label{sec:ack}
We thank Adele Reinking and Glen Liston for assisting in accessing and downscaling the MERRA dataset through MicroMet. This work is supported by the National Science Foundation (NSF) grant OPP-1839185. The opinions expressed in this work are solely those of the author(s) and do not necessarily reflect the views of the NSF.

\bibliographystyle{IEEEbib}
\bibliography{refs.bib}

\end{document}